# Enhanced hydrogen-gas permeation through rippled graphene

Wenqi Xiong[1,#], Weiqing Zhou[1,#], Pengzhan Sun[2,*], Shengjun Yuan[1,3,*]

[1]*Key Laboratory of Artificial Micro- and Nano-structures of the Ministry of Education and School of Physics and Technology, Wuhan University, Wuhan 430072, China*

[2]*Joint Key Laboratory of the Ministry of Education, Institute of Applied Physics and Materials Engineering, University of Macau, Avenida da Universidade, Taipa, Macau 999078, China*

[3]*Wuhan Institute of Quantum Technology, Wuhan 430206, China*

**ABSTRACT**

The penetration of atomic hydrogen through defect-free graphene was generally predicted to have a barrier of at least several eV, which is much higher than the 1 eV barrier measured for hydrogen-gas permeation through pristine graphene membranes. Herein, our density functional theory calculations show that ripples, which are ubiquitous in atomically thin crystals and mostly overlooked in the previous simulations, can significantly reduce the barriers for all steps constituting the mechanism of hydrogen-gas permeation through graphene membranes, including dissociation of hydrogen molecules, reconstruction of the dissociated hydrogen atoms and their flipping across graphene. Especially, the flipping barrier of hydrogen atoms from a cluster configuration is found to decrease rapidly down to <1 eV with increasing ripples' curvature. The estimated hydrogen permeation rates by fully considering the distribution of ripples with all realistic curvatures and the major reaction steps that occurred on them are quite close to the experimental measurements. Our work provides insights into the fundamental understanding of hydrogen-gas permeation through graphene membranes and emphasizes the importance of nanoscale non-flatness (ripples) in explaining many surface and transport phenomena (for example, functionalization, corrosion and separation) in graphene and other two-dimensional materials.

---

[#]These authors contributed equally to this work.

*E-mail addresses*: pengzhansun@um.edu.mo (P.Z. Sun); s.yuan@whu.edu.cn (S.J. Yuan).

# I. INTRODUCTION

It is widely believed that defect-free graphene monolayers are completely impermeable to all gas atoms and molecules [1–10]. This statement has been supported by theory predicting that even the penetration of the smallest gas atoms, helium, requires overcoming an energy barrier of at least 4 eV [1], which makes the detection of gas permeation through graphene membranes highly unrealistic under ambient conditions. Indeed, at a detection limit of $10^5$-$10^6$ atoms per second, no gas permeability could be observed through pristine graphene membranes [2]. However, thanks to the improvement in detection accuracy, a recent experimental finding has found the above conclusion to be wrong [11]. With an accuracy increased by 8-9 orders of magnitude with respect to earlier experiments, it was observed that molecular hydrogen could penetrate through pristine graphene, despite the latter being impermeable to the smaller and generally more permeable helium atoms [11]. This unexpected observation was explained by a two-step mechanism that involves the dissociation of molecular hydrogen on the graphene's surface and the flipping of hydrogen adatoms to the other side of the membrane. The first dissociation step is supported by theory suggesting that highly curved and strained graphene regions (nanoscale ripples) could dissociate molecular hydrogen and by experiments virtualizing such graphene ripples using high-resolution transmission and tunnelling microscopies [11–17]. The second flipping step was initially proposed based on the facts that the observed hydrogen-gas permeation exhibited the same 1-eV barrier as that measured for proton transport and also [18,19], a hydrogen atom being adsorbed on graphene is indistinguishable from an adsorbed proton because the latter could easily capture an electron from graphene. Nonetheless, understanding the flipping of hydrogen atoms or protons across graphene remains challenging from a theory standpoint because the predicted energy barriers are generally much higher than the experimentally measured 1 eV [1,3–7]. For example, the penetration barrier of an individual hydrogen atom through the lattice of graphene is calculated to be 2.6 to 4.6 eV [4–6]. Even for the subatomic proton, which is much smaller in size with respect to atomic hydrogen, the predicted barrier is still high and about half that of atomic hydrogen (1.2 to 2.2 eV) [4–7]. This clear inconsistency between experiment and theory was partially tackled by taking high coverage and clustering of hydrogen atoms and protons into account [8–10]. Indeed, simulations have shown that multiple hydrogen atoms and protons clustered in a localized region of the graphene membrane could decrease the flipping barrier to about 1 eV [8–10]. Despite these few theoretical efforts (mostly done using flat graphene models), it remains unknown whether nanoscale non-flatness or ripples that are ubiquitous in atomically thin crystals are important or not in understanding the observed hydrogen-gas permeation through graphene membranes, especially the proposed flipping process.

In this paper, we create nanoscale ripples in graphene by applying compressive strain and use them to simulate the realistic non-flat morphology induced by [as described in Fig. S1] [20], for example, thermal fluctuations. The importance of created local curvature is assessed in major elementary steps involved in the proposed mechanism of hydrogen-gas permeation through graphene, including dissociation of hydrogen molecules, reconstruction of the resulting hydrogen adatoms (involving diffusion and/or desorption followed by re-adsorption), their flipping and recombination [Fig. 1(a)]. We find the presence of ripples could significantly lower the energy barriers of all steps including the flipping of hydrogen adatoms. The latter process exhibits a long-lasting inconsistency between theory and experimental measurements. By fully considering the energies associated with the major reaction steps that happened on individual ripples and the latter's curvature and density distributions over a realistic large-scale graphene membrane, the estimated hydrogen permeation rates are shown to fit well with experimental measurements.

## II. CALCULATION METHODS

The theoretical calculations presented in this paper were performed using DFT implemented in the VASP package [21]. The exchange-correlation potential and ion-electron interactions were described within the generalized gradient approximation (GGA) and projected augmented wave (PAW) method [22,23]. During structural relaxation, kinetic energy cutoff and the k-point meshes were set to 500 eV and 3×3×1, respectively [24]. The van der Waals interactions were included in the hydrogen gas permeation process and treated by the semi-empirical DFT-D2 method [25,26]. All atoms were allowed to be fully optimized to the ground state, considering spin-polarization. The lattice constant of graphene was optimized as 2.46 Å. Graphene ripples were formed by applying biaxial compressive strain, and as shown in Fig. S1 [20], a stronger strain generally results in a larger curvature.

The time-dependent molecular dynamics simulation for graphene ripples was carried out by LAMMPS package [27,28]. The modified Tersoff and Airebo potentials were used to describe C-C and C-H interactions [29–31], respectively. A modelled graphene membrane contains 12800 carbon atoms, with 200 hydrogen atoms uniformly adsorbed at the central position. Periodic boundary conditions were employed to simulate an infinite structure. Their entire kinetic evolution was performed under the NVT ensemble with 5000000 steps and 1 fs per step.

We used climbing-image nudged elastic band (CI-NEB) methods to search the TS within five to eight images between IS and FS [32–35]. The TS corresponds to the state with the highest total energy during the reaction process.

## III. RESULTS AND DISCUSSION
### A. Non-flatness of graphene membranes

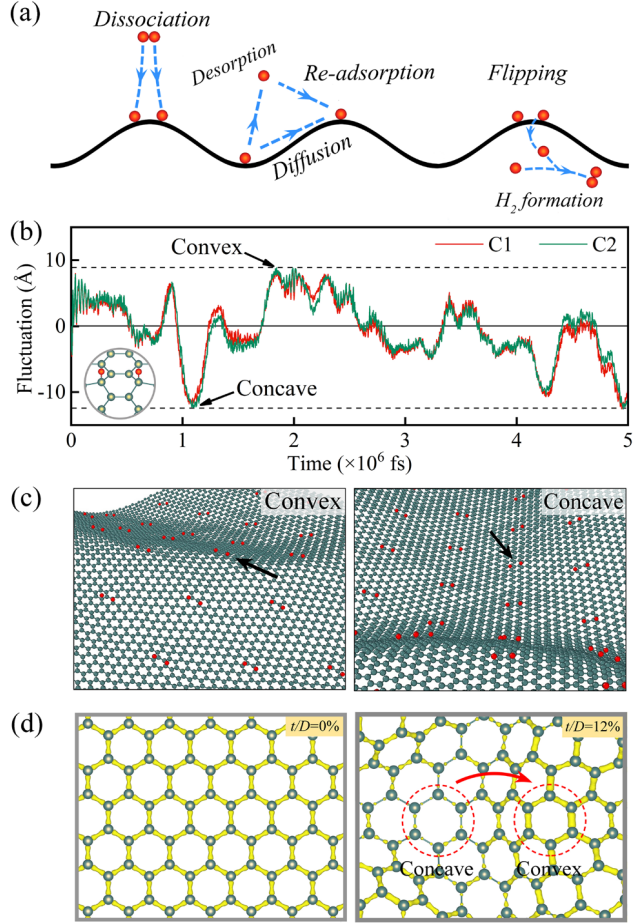

FIG. 1. (a) A schematic illustrating the major elementary steps constituting the mechanism of hydrogen permeation through graphene membranes, if local curvature is considered. (b) Time evolution of position changes for the selected carbon atoms (C1 and C2, color coded) due to thermal fluctuations. C1 and C2 are carbon atoms at the central positions and occupied by hydrogen, as illustrated in the inset. (c) Typical snapshots for the convex and concave regions of a freestanding graphene membrane captured at 300 K. The membrane is deliberately adsorbed with hydrogen adatoms. The black arrows point to atoms C1 and C2 in (b) whose vertical positions change over time. (d) Distribution of charge density for flat (left) and rippled (right) graphene. Isosurface, 0.3 $e/Bohr^3$. The arrow refers to the direction of electron transfer, from concave to convex.

In previous simulations for the penetration of atomic (hydrogen atoms) or subatomic (protons) species across graphene [1,4–10], a flat lattice model was generally adopted. This helps to simplify the simulation and reduce the cost of computation. However, this model is inconsistent with the practical freestanding graphene membranes, which are never flat and prone to nanoscale rippling due to thermal fluctuations and unavoidable local strain caused by, for example, contaminations or surface adsorbates. To take the nanoscale rippling into account, we have used classical molecular dynamics simulations to monitor time-dependent morphology changes of a freestanding graphene membrane at a finite temperature $T$ [Figs. 1(b) and 1(c)]. The thermally induced ripples can change their profiles dynamically and from concave to convex [Figs. 1(b) and 1(c)]. A redistribution of charge density between adjacent concave and convex surfaces is revealed by our density functional theory (DFT) calculations [Fig. 1(d)]. On a flat graphene lattice, the out-of-plane and unpaired electrons distribute uniformly over the entire surface [Fig. 1(d)]. Upon rippling, electrons become accumulating and depleting over the convex and concave regions, respectively. The local redistribution of charge densities could lead to changes in the energy barrier associated with each elementary step constituting the proposed mechanism of hydrogen permeation, through concave and convex regions of a graphene membrane (see below).

In the following calculations, we focus on the energy evolutions of $H_2$ dissociation, reconstruction of H atoms, and their flipping and recombination over an individual graphene ripple with a certain curvature, either convex or concave.

**B. Dissociation of hydrogen molecules**

Out-of-plane graphene ripples were created by applying compressive strain on an 8×8 supercell containing 128 carbon atoms. The curvature of a created ripple is characterized by the ratio of its height $t$ to the corrugation diameter $D$ [inset of Fig. 2(b)]. On flat and concave graphene surfaces, hydrogen dissociation is energetically unfavourable [Figs. 2(a) and S2]. The dissociation reaction is endothermic (i.e., the chemisorption energy $E_c > 0$) and the barrier $E_b$ is at least 3 eV. This makes the dissociation of hydrogen on flat and concave graphene highly improbable. On the other hand, the presence of convex ripples with considerable curvature $t/D$ makes the dissociation reaction energetically favourable [Fig. 2(a)], in agreement with recent simulations [11,16,36]. For example, if we consider a graphene ripple with curvature $t/D =$ 12%, the dissociation of molecular hydrogen on the central positions [inset of Fig. 1(b)] is calculated to be exothermic ($E_c < 0$) with an energy barrier $E_b$ of only about 0.62 eV. Both $E_c$ and $E_b$ are found to decrease monotonically with increasing $t/D$ [Fig. 2(b)]. The critical curvature $t/D^*$ at which the dissociation reaction becomes exothermic is about 10%. Ripples with $t/D$ of larger than this critical value result in a dissociation barrier of <1 eV. The presence of such large ripples ensures that the dissociation of molecular hydrogen is not a rate-limited step as compared with the flipping of the resulting hydrogen adatoms (protons adsorbed on graphene). The latter step was shown to have an energy barrier of about 1 eV, according to the proton transport measurements [18,19]. Hydrogen dissociation in other positions, including the bridge position, were also considered, but we found they are all energetically less favourable as compared with the central position [Supplementary Note1 and Figs. S3 and S4] [20].

Dissociation of molecular hydrogen on the graphene's surface critically involves the cleavage of H-H bonds in the split $H_2$ and the formation of new C-H bonds on the graphene surface. Accordingly, the energy evolution should be determined by the changes in length of the involved H-H and C-H bonds. Trying to understand the correlation between energy and bond length, we have plotted the potential energy surface (PES) as a function of the distance between two H atoms in the dissociated molecule ($d_{H-H}$) and their distance to the central positions ($d_{H-C}$) of both flat and rippled graphene surfaces [Figs. 2(c) and 2(d)]. Comparing with flat graphene [Fig. 2(c)], $H_2$ dissociation on rippled graphene requires a shorter elongation of H-H bonds at the transition state (TS) and simultaneously, the C-H formation can happen at a longer distance [Fig. 2(d)]. The overall effect is a notably smaller energy cost for dissociation on rippled graphene than that on flat graphene, as evidenced in PES [Figs. 2(c) and 2(d)]. Of the described bond length changes, the H-H split is expected to contribute to most of the barrier at the TS. Indeed, our calculations in Fig. 2(e) have shown that the energy change is highly sensitive to $d_{H-H}$ of the dissociated $H_2$ molecule and the energy gain upon elongation of H-H bonds at the TS ($E_b^{H-H}$) is close in magnitude to the total $E_b$ for the full range of curvature considered ($t/D$, 0%~12%) [Fig. 2(f)]. To quantitatively disentangle the contribution of changes in bond lengths ($d_{H-H}$ and $d_{H-C}$) to $E_b$, we have performed the following analysis. We first calculated $d_{H-H}$ at the TS when $H_2$ underwent dissociation over graphene ripples with different curvatures $t/D$, and then plotted them together with $E_b^{H-H}$ [from Fig. 2(e)] and $E_b$ [from Fig. 2(b)] as a function of $t/D$. As shown in Fig. 2(f), the function of $d_{H-H}(t/D)$ correlates well with those of $E_b^{H-H}(t/D)$ and $E_b(t/D)$ over the entire range of $t/D$ considered. Small differences between $E_b^{H-H}$ and $E_b$ at a certain $t/D$ can be attributed to C-H formation, which also requires overcoming a finite energy barrier albeit small [Fig. 3(c)]. More analyses can be found in Supplementary Note2.

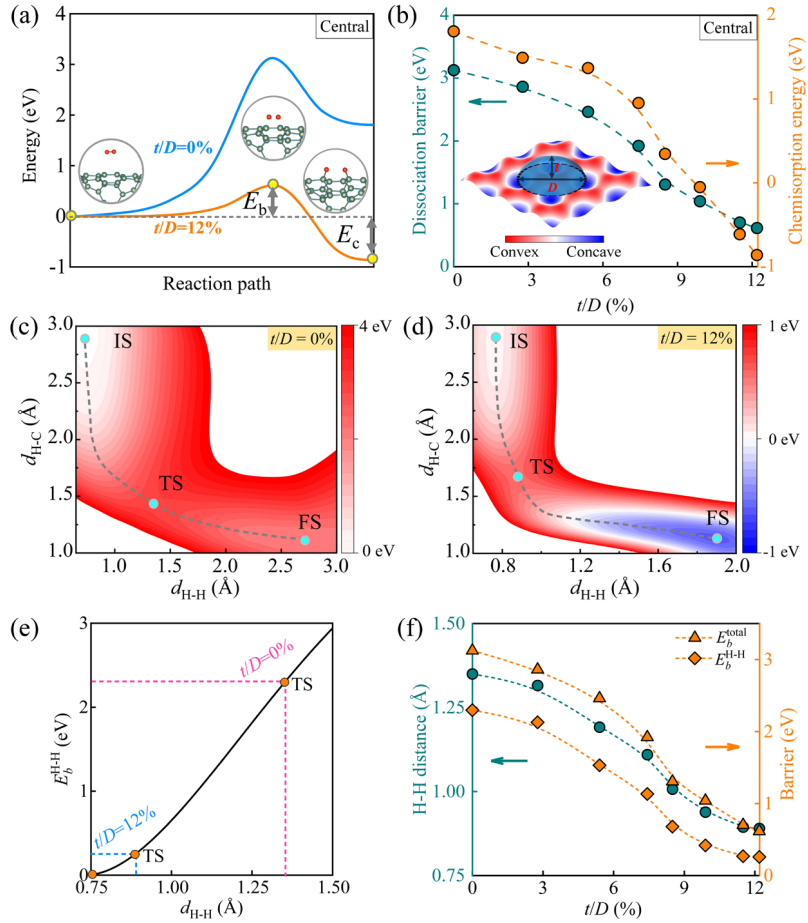

FIG. 2. (a) Dissociation of $H_2$ at the central positions of flat and rippled ($t/D$=12%) graphene (color coded). Insets: atomic configurations of initial (IS), transition (TS), and final (FS) states, respectively. Dissociation barrier $E_b$ and chemisorption energy $E_c$ are illustrated on the energy profile. (b) $E_b$ (left y-axis) and $E_c$ (right) as a function of the ripple's curvature $t/D$ (illustrated in the inset) for dissociation of $H_2$ at the central position of graphene. Inset: the ripple's curvature is characterized by the ratio $t/D$ of its height $t$ to diameter $D$. Positive (negative) values represent a convex (concave) shape. Dashed lines are guides to the eyes. The potential energy surface (PES) for dissociation of $H_2$ on (c) flat and (d) rippled ($t/D$=12%) graphene as a function of the distance between two H atoms in the dissociated molecule ($d_{H-H}$) and their distance to the central adsorption sites of the lattice ($d_{H-C}$). The dashed lines are the minimum energy paths along the reaction path. (e) The energy gain of $H_2$, $E_b^{H-H}$, with increasing $d_{H-H}$ from its equilibrium value of 0.75 Å. The bond length measured at TS for $t/D$ = 0 and 12% are labelled. (f) At different curvatures $t/D$, $d_{H-H}$ measured at TS, and the resulting $E_b^{H-H}$ obtained from (c) are compared with the total energy barrier $E_b$ obtained from (b). Dashed lines are guides to the eyes.

## C. Reconstruction of hydrogen adatoms

After dissociation on a convex ripple, the resulting hydrogen adatoms are expected to form strong C-H bonds with the lattice carbon atoms due to the exothermic dissociation reaction with a high and negative chemisorption energy [for example, l$E_c$l is close to 1 eV at $t/D$ = 12%, Fig. 2(a)]. On the other hand, the C-H formation should be weaker on a concave or flat graphene surface because hydrogen dissociation on such surfaces is energetically unfavourable having $E_c$ >0 [Fig. 2]. This is in conceptual agreement with a few simulations reporting the modulation of adsorption and desorption energies of hydrogen atoms on graphene by local curvatures [37,38]. The dynamic morphology transition of graphene ripples from convex to concave could be helped by, for example, thermal fluctuations [Fig. 1]. Therefore, one would expect the hydrogen adatoms to either desorb from the surface or undergo local reconstruction via desorption followed by re-adsorption to form H clusters when the graphene ripples evolve from convex to concave. The local clustering structures are predicted to be energetically more stable with respect to hydrogen atoms being randomly adsorbed on the graphitic surfaces and have been observed using tunnelling microscopy [38–41].

To assess the importance of ripples (either convex or concave) on the formation of H clusters via desorption and re-adsorption, we need to know the desorption and re-adsorption barriers on graphene ripples having different curvatures $t/D$ [Fig. 3]. On a flat graphene surface, the desorption barrier is about 1 eV, in agreement with the previous simulations (0.7-1.1 eV) [39,41]. Ripples with a concave profile significantly reduce the barrier required for desorption. For example, the desorption barrier for a concave surface with $t/D$ = 12% is only 0.56 eV [Figs. 3(a) and 3(b)]. In contrast, ripples having the same $t/D$ but a convex profile increase the barrier to as high as 2.46 eV [Figs. 3(a), and S5 for energy profiles of desorption/diffusion]. If we define the curvatures of concave and convex ripples as negative and positive, respectively, we would find the desorption barrier decreases monotonically with decreasing $t/D$ and the barrier changes more rapidly with the curvature on convex ripples than that on concave ones [Fig. 3(a)]. This is contrary to the dissociation of molecular hydrogen [Fig. 2] which critically involves adsorption of hydrogen adatoms, the inverse process of desorption. Furthermore, it is worth noting that the desorption of hydrogen adatoms is slightly endothermic even for a concave ripple with high curvature. For example, the final energy at $t/D$ = -12% is about 0.1 eV [Fig. 3(b)]. Such a small energy gain could, however, be easily compensated via the energy released from the dissociation step [Fig. 2] or direct heat treatment. Indeed, experiments have observed the heating-induced desorption of hydrogen atoms from a graphite surface, even if the barrier is predicted to be much higher than that for the concave graphene ripples considered here [Fig. 3(a)] [39,41].

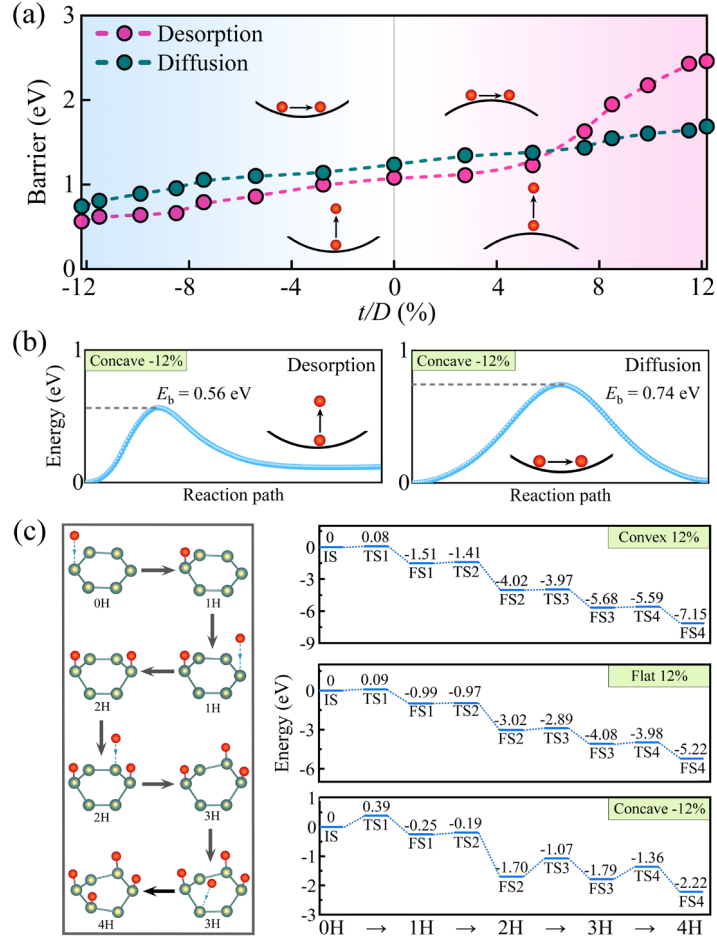

FIG. 3. (a) Desorption and diffusion barriers of hydrogen adatoms on convex, flat and concave surfaces. Insets: schematics for the desorption and diffusion processes on concave and convex surfaces. (b) Energy profiles for the desorption (left) and diffusion (right) of H atoms over a concave ripple ($t/D$ = -12%). Insets illustrate the desorption and diffusion processes with the arrows showing the moving direction of H atoms. (c) Left panel: schematic diagram of the sequential re-adsorption of four hydrogen atoms. Blue dashed arrows indicate the moving direction of H atoms. Right panel: energy evolution of the re-adsorption reaction for convex ($t/D$ = +12%), flat and concave ($t/D$ = -12%) surfaces (from top to bottom). The term '$n$H' ($n$ = 1 to 4) denotes the number of H atoms being adsorbed on the hexatomic ring.

As an inverse process of desorption, the re-adsorption of hydrogen atoms onto the surface of graphene should exhibit an opposite trend of energy evolution with its curvature. Indeed, we have performed calculations for the sequential addition of hydrogen atoms (up to 4) onto a hexatomic ring within a graphene ripple and found that re-adsorption on a convex surface is more energetically favorable than that on a concave surface [Fig. 3(c)]. For all curvatures calculated, the adsorption of multiple hydrogen atoms results in a gradually lowered energy [Fig. 3(c)], that is, clustering of H atoms is energetically favored, in agreement with the previous theoretical and experimental reports [39–41]. All the calculated barriers for re-adsorption are minor with respect to desorption, regardless of ripple's local curvatures.

In addition to the described desorption and re-adsorption, another possible way of forming local H clusters could be the diffusion of chemisorbed H atoms. However, the barrier for diffusion along a flat graphene or graphite surface is predicted to be higher than (>1.1 eV) than that for desorption under similar conditions [39]. Same results were also obtained in our simulations [Fig. 3(a)]. Compared with desorption, the barrier for diffusion along a rippled graphene surface decreases steadily but more slowly from ~1.7 eV to ~0.7 eV with decreasing the curvature in the same range from +12% to -12% [Fig. 3(a)]. Furthermore, we have examined the formation of a 4H cluster within the same hexatomic ring through dissociation followed by sequential adsorption of two $H_2$ molecules. Our results showed that the dissociative adsorption of the second $H_2$ is an endothermic process with $E_c$ > 1.8 eV. These calculations indicate that comparing with desorption followed by re-adsorption of atomic H, the diffusion of H adatoms and dissociative adsorption of multiple $H_2$ are less probable in terms of the reconstruction of hydrogen atoms on graphene's surface and hence, the formation of H clusters. The latter local configurations are important for the flipping step (see below).

### D. Flipping of hydrogen atoms

Direct penetration of individual hydrogen atoms through a flat graphene lattice is predicted to be extremely difficult under ambient conditions with an energy barrier of at least several eV [4–6]. This is consistent with our simulations showing a flipping barrier $E_f$ of >4 eV for a chemisorbed hydrogen adatom [Fig. S6 for flipping of flat and rippled] [20]. Even if the presence of ripples is considered, the resulting $E_f$ still shows a negligible variation with respect to the flat case. On the other hand, the simulations in Fig. 3(c) indicate that hydrogen adatoms tend to cluster on the graphene's surface. The formation of such H clusters is energetically favourable [Fig. 3(c)] and has been observed experimentally [38,39]. Recent simulations suggest that the clustering of hydrogen atoms and protons could help lower the flipping barrier $E_f$ [8–10]. This effect is also taken into account in our following simulations [Fig. 4].

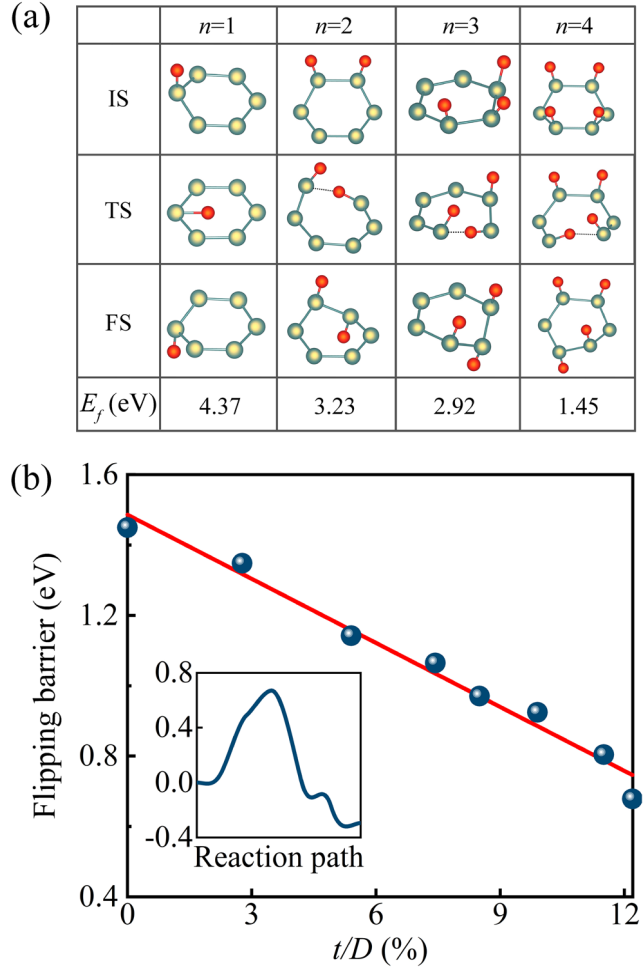

FIG. 4. (a) Examples of the local cluster configurations with number of hydrogen atoms $n$ = 1-4 and corresponding flipping barriers. (b) The flipping barriers for the four-hydrogen configuration at different curvatures. Symbols: calculated data. Solid line: best linear fit. Inset: energy profile at $t/D$ = 12%.

To seek for the most energetically favorable clustering configuration, we have performed systematic calculations for the flipping barriers across a flat graphene lattice by sequentially adding hydrogen atoms onto a hexatomic ring (total number $n$ of H atoms, 1-6), similar to that done in Fig. 3(c). Examples for selected configurations with 1-4 of H atoms are shown in Fig. 4(a). More configurations and analyses are provided in Table S1 and Fig. S7 for different clusters [20]. Generally, the flipping barrier $E_f$ across a flat graphene lattice decreases with increasing the number of H atoms forming the cluster [Figs. 4(a) and S7] [20]. The scatters of $E_f$ calculated for the same size clusters can be as large as 3 eV and are highly sensitive to the detailed arrangement of hydrogen atoms [Table S1 and Fig. S7 for different clusters] [20]. After trying many different configurations, we have found flipping from a cluster containing four hydrogen adatoms being the most energetically favourable [Fig. 4(a)], in qualitative agreement with the previous simulations [10]. The flipping process starts from two pairs of hydrogen adatoms occupying two of the three sets of central positions in the hexatomic ring [IS for $n$ = 4 in Fig. 4(a)], forming four sp$^3$ C-H hybridizations. Then one of the bonded hydrogen atoms flips inward and breaks the C-C bond connecting two sp$^3$ hybridized carbon atoms, resulting in a transient sp$^2$ C-H pair [TS for $n$ = 4 in Fig. 4(a)]. Finally, the flipped hydrogen atom continuously rotates to the other side of the lattice and simultaneously, a new C-C bond is established between the sp$^2$ C-H pair, turning them into the more stable sp$^3$ hybridizations [FS for $n$ = 4 in Fig. 4(a)]. The flipping barrier $E_f$ at TS is calculated to be ~1.5 eV in this configuration for a flat lattice. Such a low $E_f$ is mostly attributed to an enhanced bonding strength with respect to other configurations, according to the crystal orbital Hamilton population (COHP) bonding analysis in Supplementary Note3 and Fig. S8 [20].

Next, we have used the described 4H configuration to calculate the flipping barriers $E_f$ at different curvatures $t/D$. Different from the flipping of individual hydrogen atoms [Fig. S6 for flipping of flat and rippled] where the presence of ripples induces a negligible change in $E_f$, the flipping barrier of hydrogen atoms from the 4H cluster decreases rapidly and linearly with the increase of $t/D$ [Fig. 4(b)] [20]. At the maximum curvature considered, $t/D$ = 12%, the minimum $E_f$ is calculated to be ~0.7 eV. Other cluster configurations including the one containing 6 hydrogen atoms as well as concave ripples were also considered in Fig. S9 for both 4H and 6H with concave [20], but we found they are less energetically favourable compared with the described 4H configuration. Furthermore, after flipping from the convex to the concave side of the membrane, the recombination of hydrogen adatoms from the concave surface is shown to be energetically more favourable than the flipping step [Fig. S10 for recombination] [20].

Altogether, compared with the previously predicted high barriers (2.6 to 4.6 eV) for the penetration of individual hydrogen atoms across flat graphene [4–6], the considerably lowered $E_f$ obtained by combined clustering of hydrogen atoms and graphene's rippled morphology makes the flipping of hydrogen atoms across defect-free graphene membranes plausible under ambient conditions.

**E. Hydrogen-gas permeation through graphene membranes**

The above calculations are mainly done with individual ripples. It would be more practical to consider a graphene membrane [for example, the ones in Fig. 1 and inset of Fig. 5(a)], which contains numerous ripples with a wide size and density distribution. To this end, we have estimated the hydrogen-gas permeation through a freestanding graphene membrane by considering the above-described elementary steps happened on ripples with all possible curvatures that can be supported by the membrane without breaking it (for example, $t/D$ <15%).

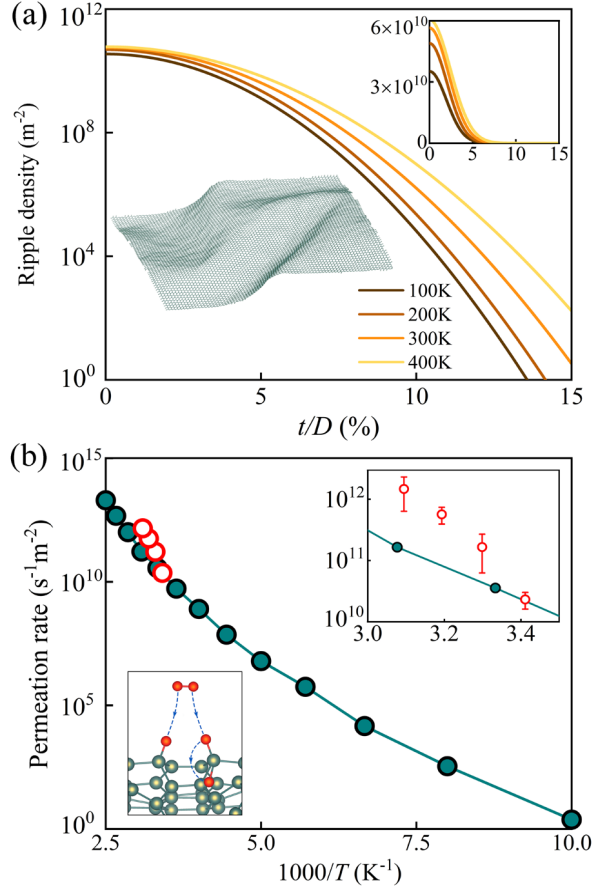

FIG. 5. (a) Ripple density as a function of curvature $t/D$ at different $T$ (color coded). The data are cited from reference [42]. Upper inset: same data as in the main panel but in a linear y scale. Lower inset: a typical snapshot of a graphene membrane at 300 K from molecular dynamics simulations. (b) Calculated hydrogen permeation rates at different $T$ by considering the ripple distribution in (a) and assuming flipping as the main rate-limited step. Green symbols: calculated data at different $T$. Red symbols: experimental data cited from reference [11]. Upper inset: zoom-in for the experimentally relevant $T$ range. Lower inset: schematics for the hydrogen permeation process including dissociation and flipping.

The distribution of ripple densities $P(t/D)$ is cited from our recent report and shown in Fig. 5(a) [42]. The ripple density is seen to follow the Boltzmann distribution [inset of Fig. 5(a)] and decays rapidly with increasing curvature. A higher temperature inducing stronger thermal fluctuations results in larger ripple densities with high curvatures [Fig. 5(a)]. We have taken the simulated $P(t/D)$ into account in the following estimation of hydrogen-gas permeation rates $\Gamma$. To simplify the calculation, we recall that only on ripples with curvature $t/D$ >10%, the dissociation of molecular hydrogen is energetically favourable with the chemisorption energy $E_c$ <0 and dissociation barrier $E_b$ <1 eV [Fig. 2(b)] and hence, in the following estimations, we are only interested in these large ripples. For such large ripples, the minimum flipping barriers $E_f$ calculated for the most energetically favourable four-hydrogen cluster [Fig. 4] are mostly higher than the dissociation barrier $E_b$ at a large $t/D$ [Fig. S11] [20]. Other cluster configurations exhibit even higher $E_f$ [Fig. S7 for both 4H and 6H with concave] [20]. Therefore, the flipping step is considered as the main rate-limiting step in our calculations, which is consistent with the previous report [11]. Note that graphene ripples can change their profiles from convex to concave dynamically and back [Fig. 1], and also, the proposed four-hydrogen cluster could, in principle, be formed by the re-adsorption of additional hydrogen atoms that desorb from adjacent regions, for example, concave surfaces [Fig. 3].

The flipping frequency $v$ of chemisorbed hydrogen adatoms across the graphene lattice can be evaluated as $v = k_B T/h \left[ \exp(-E_f/k_B T) \right]$, where $k_B$ and $h$ are Boltzmann and Planck constants, and $k_B T/h$ describes the vibration frequency of bonded hydrogen atoms. In principle, when calculating the activation energy of chemical reactions involving hydrogen, the zero-point energy (ZPE) effect needs to be considered. We have also taken the ZPE effect into account in our calculations but found that after ZPE correction, the activation energy, especially for the flipping step, exhibits little difference with respect to $E_f$ [Table. S2]. Multiplying the flipping frequency $v$ by the ripple density [Fig. 5(a)] and integrating over the range of large curvatures (10%< $t/D$ <15%), the hydrogen-gas permeation rates $\Gamma$ through a graphene membrane are estimated and plotted in Fig. 5(b). Detailed calculations are provided in Supplemental Materials [20]. In general, the estimated $\Gamma$ using our theoretical models are quite close to the experimental measurements [11]. The maximum deviation appears at the highest temperature used in experiments ($T \approx 60$ °C) and is still no larger than one order of magnitude. The slight underestimation of the permeation rates indicates a lower theoretical activation energy was used with respect to experimental measurements (1.0 ± 0.1 eV). This can be attributed to the fact that in addition to the most energetically favourable four-hydrogen configuration, other H clusters could also be present, which impose a higher barrier to the rate-limiting flipping step [Fig. S7 for both 4H and 6H with concave] [20].

In the above described mechanism for hydrogen-gas permeation through graphene membranes, it would be of interest to compare the ripples' lifespan with the timescale of the dissociation-flipping process of H₂. Our molecular dynamics simulations revealed multiple convex-concave transitions over the total time period of $10^6$ fs [as shown in Fig. 1(b)], with the ripple's lifespan being at least 100 picoseconds. On the other hand, the timescale $\tau$ of the dissociation-flipping process can be estimated from the chemisorption energy $E_c$ as $\tau = h/E_c$ where $h$ is Planck's constant. Taking the chemisorption of hydrogen atom on graphene as a reference, the adsorption energy is about 0.8-1.2 eV and the timescale $\tau$ is on the order of femtoseconds [39,41]. Clearly, the ripple's lifetime is several orders of magnitude longer than the typical timescale of the dissociation-flipping process, allowing for the occurrence of hydrogen permeation. Furthermore, it would also be interesting to consider the nuclear quantum effects for systems involving hydrogen like in the case here [43], which have a significant impact on chemical reactions. We have taken into account these effects by evaluating isotropic effects using DFT calculations. After applying the ZPE correction to IS, TS, and FS, we have found that the dissociation barriers of deuterium and tritium were similar to those of hydrogen, with only a slight difference of about 1 meV. However, due to the heavier nuclei, deuterium and tritium are expected to have slower rates of dissociation, flipping, and permeation.

In this study, we investigated not only static ripples but also dynamic ripples under thermal fluctuation. Under these conditions, the graphene membrane undergoes lattice distortion, causing in-plane strain. As seen in the statistical distribution presented in Fig. S12(a), the C-C bond can experience a maximum tensile strain of up to 10%. Additionally, in Fig. S12(b), we calculated the dissociation barriers of $H_2$ on graphene under varying degrees of tensile strain and found that they remained high (>2.5 eV) even when the strain exceeded 10%. This suggests that curvature plays a more critical role than in-plane strain in chemisorption processes, which is consistent with our previous experimental findings [11].

## IV. CONCLUSIONS

A flat sheet of graphene imposes very high barriers of at least several eV for the dissociation of molecular hydrogen and the cross-lattice flipping of hydrogen atoms according to theory. This makes the experimental observation of hydrogen-gas permeation through pristine graphene membranes highly elusive. On the other hand, ripples that are unavoidable in atomically thin crystals significantly lower the energy barriers required by the major steps involved in the proposed mechanism for the observed hydrogen permeation, including dissociation, reconstruction and flipping. Combining the effects of ripples and clustering structures of H adatoms on graphene, the energy barrier for the flipping step, which is believed to be rate-limiting from both experimental and theoretical standpoints, can be effectively lowered down to <1 eV. Bearing these considerations in mind, the estimated hydrogen gas permeation rates through a freestanding graphene membrane where ripples with all possible curvatures are present are in good agreement with the experimental measurements. Our work provides implications for a fundamental understanding of hydrogen permeation through defect-free graphene and emphasizes the importance of non-flatness (ripples), especially in explaining the flipping step. The presence of ripples is mostly overlooked in calculations of the transport of molecules and ions, including protons through graphene membranes [1,4–10,44–46], which would also be important for the modification and chemical functionalization of graphene and other two-dimensional materials [47–49]. Very recently, we have successfully explained the phenomenon of the impermeability of hydrogen gas through rippled h-BN using this theoretical mechanism [50], which also provides experimental support for our work.

## ACKNOWLEDGEMENTS

This work was supported by the National Natural Science Foundation of China (Grants No. 12174291), the National Key R&D Program of China (Grant No. 2018YFA0305800), the fellowship of China Postdoctoral Science Foundation (Grant No. 2021M702532) and the Natural Science Foundation of Hubei Province of China (Grant No. 2020CFA041). All numerical calculations presented in this paper were performed on the supercomputing system in the Supercomputing Center of Wuhan University.